# Inclusion unlocks the creative potential of gender diversity in teams

*Balázs Vedres*
Central European University, University of Oxford

*Orsolya Vásárhelyi*
Laboratory for Networks, Technology and Innovation Corvinus Institute for Advanced Studies, Corvinus University of Budapest

**Abstract**

Diversity in teams can boost creativity, and gender diversity was shown to be a contributor to collective creativity. We show that gender diversity requires inclusion to lead to benefits in creativity by analyzing teams in 4011 video game projects. Recording data on the weighted network from past collaborations, we developed four measures of inclusion, depending on a lack of segregation, strong ties across genders, and the incorporation of women into the core of the team's network. We found that gender diversity without inclusion does not contribute to creativity, while with maximal inclusion one standard deviation change in diversity results in .04 to .09 standard deviation change in creativity, depending on the measure of inclusion. To reap creative benefits of diversity, developer firms need to include 23% or more female developers (as opposed to the 15% mean female proportion) and include them in the team along all dimensions. Inclusion at low diversity has a negative effect. By analyzing the sequences of diversity and inclusion across games within firms, we found that adding diversity first, and developing inclusion later can lead to higher diversity and inclusion, compared to adding female developers with already existing cross-gender ties to the team.

**Introduction**

Groups with diverse members can be engines of creativity. Project teams – small collectives recruited for a defined task – are often used to address creative tasks, where a nonroutine solution is needed to solve a problem (Amabile 1983; Kozlowski and Bell 2013), and such teams boost returns to resources in innovative organizations (Cohen and Bailey 1997; Wuchty, Jones, and Uzzi 2007). There is evidence that teams possess collective intelligence beyond the mean or maximal individual intelligence of team members (Woolley et al. 2010). It is also often demonstrated, that the collective intelligence and creative capacity of teams is a function of their cognitive diversity (Horwitz and Horwitz 2007). When team members come from diverse demographic backgrounds and have diverse past experiences, they have a higher openness to divergent thinking (Levi 2017), and they are more willing to constructively challenge the status quo (Amabile et al. 1996).

Gender diversity specifically has been shown to boost collective intelligence (Woolley et al. 2010; Xie et al. 2020), and the low proportion of non-dominant genders dampens innovative potential in teams (Beede et al. 2011; Hofstra et al. 2020). Women, transgender, and gender-nonconforming people (TGNC) are under-represented in STEM fields – especially in computer science and software careers (Cheryan et al. 2017; Haverkamp et al. 2021) –, and even if they embark on a career in technology, they are less appreciated and successful, and are more likely to leave at various key stages compared to men (Clark Blickenstaff 2005; Vedres and Vasarhelyi 2019). It is important to analyze gender diversity in STEM teams to understand how diversity contributes to innovation when females are in minority, and often face discrimination (Brooke 2019).

Despite a general agreement about the promise of diversity for creativity, studies on how team diversity leads to an increase in team performance has not reached a clear consensus (Joshi and Roh 2009; van Knippenberg, De Dreu, and Homan 2004). It is clear that group creativity is not a simple function of individual creativity, but a complex interplay of compositional diversity, internal team structures, and the organizational-cultural environment of the team (Vedres and Stark 2010; Woodman, Sawyer, and Griffin 1993). On one hand diversity itself, while contributing to openness to creative solutions, can contribute to weakened team cohesion, and heightened conflict (Webber and Donahue 2001). On the other hand, the right routines and communication structure within the team can multiply the power of diversity for innovation (Cohen and Bailey 1997). Thus we need to consider diversity together with inclusion to understand the potential of diversity for collective creativity (Milliken and Martins 1996).

To understand creativity in diverse collectives we need to pay attention to inclusion as well (De Dreu and West 2001; Ferdman and Deane 2014; Mor Barak et al. 2016). Diversity without inclusion can lead to mistrust and a breakdown of communications (Pelled 1996), preventing a true dialog where diverse approaches to the problem at hand can be explored (Nishii and Goncalo 2008; Pearsall, Ellis, and Evans 2008). The mere increase in the proportion of women in a field will not eliminate their discrimination (Begeny et al. 2020). When diverse teams are well integrated, even if diversity results in conflicts, such conflicts can be beneficial to performance in complex, non-routine tasks (Jehn 1995). We argue that in teams with a discriminated minority – the case with gender in STEM –, without inclusion diversity will not have a positive impact on collective creativity, as various perspectives that diverse participants bring to the team would not have a chance to be contrasted and utilized.

### *Gender diversity and collaboration in video game development*
We analyze teams in the video game industry from 1994 to 2009. Video games is today by far the largest entertainment industry, that overtook movies and music in terms of gross revenues in 2003, and by 2009 it became larger than movies and music combined (Kooistra 2019). Video game development is a field that prizes creativity and distinctiveness (de Vaan, Vedres, and Stark 2015), but it is a male-dominated field, where only about 17% of developers were female in 2010, and about 20% of them are female today (Bailey, Miyata, and Yoshida 2021). The content of video games is decidedly male, as only about 13% of all characters depicted in video games are female (Williams et al. 2009). Fields with low proportion of female participants feature strong prejudice and discrimination against women (Begeny et al. 2020), thus if we are able to show creative advantage to gender diversity in a strongly male dominated field, such as video games, it would serve as a strong evidence for the power of gender diversity.

We measure weighted collaborative ties between developers as the number of prior joint participations in game development projects, following others who have analyzed collaboration in co-authorship (Newman 2001), movies (Lutter 2015), musicals (Uzzi and Spiro 2005), video game development (de Vaan et al. 2015), or jazz music (Phillips 2011).

These approaches take a bipartite graph of person-to-event affiliations (affiliations to papers published, movies, games, or albums released), and analyze the person-to-person projection, an undirected weighted graph, where $w_{ij} = \sum a_{ik} a_{jk}$, if $k$ is a shared affiliation for $i$ and $j$ at time $t-1$ that predates time $t$ of the focal event analyzed.

*Measuring inclusion in diverse teams*
Inclusion in a work team context can be defined as actively engaging with team members across differences (Ferdman and Deane 2014). We cannot speak of inclusion, when different team members are excluded from meaningful contact and collaboration, either by isolating individuals that are different, or allowing the team to fragment into homophilous subgroups. Inclusion is also absent when team members are fully assimilated, and their diversity becomes muted and irrelevant in collaborations (Shore et al. 2011).

There is no consensus about how inclusion should be measured. A wide range of measures were proposed, that include dimensions of individual or group experiences, leadership, norms and values (Ferdman and Deane 2014), influence on decisions, and access to resources (Mor-Barak and Cherin 1998), a sense of belonging, and authenticity (Jansen et al. 2014), organizational support, and tolerance towards uniqueness (Chung et al. 2020). These measures require reactive data collection techniques (like surveys or interview methods), and are not scalable to large observational data.

In this article we focus on the relational aspects of inclusion, and we rely on weighted graph measures of how well gender minority members in the team are connected into the collaboration network, as evidenced by past projects. We build on past research that developed related measures conceptualized as network heterogeneity (Reagans and Zuckerman 2001), or the co-presence of incumbency and network diversity (Guimera et al. 2005). We develop a range of measures for inclusion, from a minimal level – lack of segregation along gender – to a stronger level – the presence of gender minority in the network core of the team.

We define three dimensions of inclusion. First, we define inclusion as mixing: the lack of segregation along gender. Second, we define inclusion as bonding, where inclusion is stronger if the ties that connect across gender categories are of higher weight. And third, we define inclusion as incorporation when gender minorities are included in the core of the team's network. Figure 1 illustrates these measures examples of weighted collaboration graphs, where the number of nodes and the proportion of genders are kept constant.

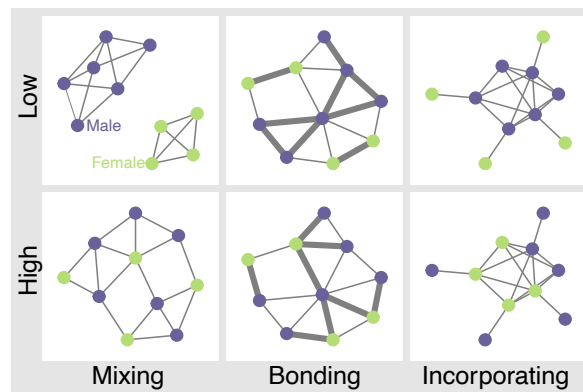

**Figure 1:** Illustrations of low and high inclusion by our three measures.

Our first measure of inclusion is *mixing*: the lack of segmentation by attributes in the team. If team members are segregated, the team cannot benefit from exchanges across gender lines. Limiting collaborative connections that would allow for negotiating diverse perspectives, essential for innovation (Northcraft et al. 1995). When gender underlie subgroup formation, team identity erodes, and the salience of gender identity increases often to the extent of gender conflict (Carton and Cummings 2012). Network fragmentation along attributes can be measured by assortativity, the over-representation of ties within categories (Newman 2002).

Our second measure of inclusion is *bonding*, that captures the strength of ties across gender categories. Strong ties are seen to be vehicles of trust (Granovetter 1973), and they offer high-bandwidth interpersonal channels that are crucial in innovative contexts, when the information environment is complex, and updates frequently (Aral and Van Alstyne 2011). The stronger the ties in mixed gender dyads, the broader the social bandwidth a team can rely on to develop novel solutions.

Our third measure of inclusion is *incorporating*: the proportion of female team members in the core of the team's network. Being in the core opens access to informal leadership, and thus offers the opportunity for women to have a say in decision making (Shore et al. 2011). Women in leadership positions tend to encourage participation, and facilitate broader information sharing (Rosener 2011), and encourage innovation and risk-taking (Adams and Funk 2012). Our measure of *combined inclusion* is then the product of the three raw measures, representing co-occurrence of various forms of inclusion. (See Quantitative Measures in Materials and Methods for formulas of inclusion metrics.)

**Results**

*The impact of diversity and inclusion on creativity*
Gender diversity in video game development is low (the proportion of females is .15 overall), even less than the proportion of females in STEM and computer programming (Beckhusen 2016; United States Department of Labor 2015), that is around twenty percent. As shown on Figure 2, the female proportion of game developers had been slowly increasing from .12 in 1994 to .18 in 2009. There is no comparable increase in inclusion, as the combined inclusion index hovers around the average of .06 without a significant trend.

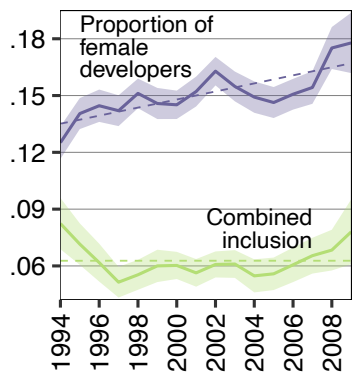

**Figure 2:** Gender Diversity and Inclusion over time. Solid lines indicate the mean with a 95% CI. Dashed lines show OLS trendline. Gender diversity is measured on the full sample of 8,617 games, while Combined Inclusion is measured on the sample of games where there was at least one female developer (4,011 games).

As expected by the literature (Bear and Woolley 2011; Beede et al. 2011; Hofstra et al. 2020; Woolley et al. 2010; Xie et al. 2020), we find that gender diversity is positively related to creativity in video game projects, as an increase in gender diversity means a slight increase in creativity, without considering inclusion. Considering games as units of analysis, one standard deviation increase in gender diversity measured by the Blau diversity index results in .09 (95% CI: .06; .12) standard deviation increase in creativity measured by game distinctiveness. At the firm level this relationship is slightly stronger, as one standard deviation increase in developer firm average gender diversity results in .13 (95% CI: .08; .18) standard deviation increase in average developer firm level distinctiveness. (See firm-level point estimates in Figure S2, panel d in Supplementary Materials.)

Figure 3. shows points estimates of standardized gender diversity, forms of inclusion, and the interaction of gender diversity and inclusion. Once we include any of the three forms of inclusion in our model (or combined inclusion as the product of our three inclusion measures), we no longer see a positive main effect for gender diversity. The signs and significance levels of these point estimates are comparable when we control for developer firm level heterogeneity (with random or fixed effects for developer firm intercepts), or when we aggregate the data to the developer firm level. In fact, the main effect for gender diversity becomes significant negative at the developer firm level (point estimate of 0.36, and 95% CI: -.68; -.05). This suggests that gender diversity without inclusion does not contribute to group creativity.

The main effect of inclusion is mostly negative: two of our three measures (mixing and bonding) show significant negative coefficients, both at the game level (mixing: -.07, 95% CI: -.12; -.01; bonding: -.13, 95% CI: -.21; -.06), and at the developer firm level as well (mixing: -.14, 95% CI: -.24; -.05; bonding: -.17, 95% CI: -.30; -.04). These estimates indicate that inclusion without gender diversity does not help creativity. Of course, inclusion is not interpretable for zero diversity; these estimates indicate that increasing inclusion for minimal levels of diversity will not help creativity. It is likely that high inclusion at low levels of diversity leads to assimilation and tokenism that was shown to nullify creative benefits to diversity (Shore et al. 2011).

Creativity in game development benefits from gender diversity *with* inclusion. Game developer teams should include female collaborators, and also integrate them with the rest of the team to boost creativity. The interaction of gender diversity and inclusion is positive and significant for all of our inclusion measures and shows a significant if moderate contribution to distinctiveness. At the game level, one standard deviation change of gender diversity and inclusion jointly results in an increase from .04 to .06 standard deviations in game distinctiveness (in addition to the main effects of diversity and inclusion). At the developer firm level, one standard deviation change of gender diversity and inclusion jointly results in increases from .04 to .09 standard deviations in firm level average game distinctiveness.

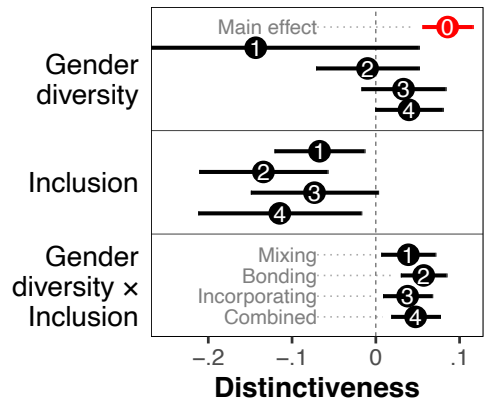

**Figure 3:** Point estimates of distinctiveness with 95% CI for gender diversity, four variables of inclusion, and their interactions with gender diversity. Markers are numbered according to OLS models; coefficients are for one SD change in distinctiveness as a result of one SD change in independent variables.

Our point estimates suggest that teams should make sure female collaborators are included into team networks to be more creative, but teams with average gender diversity (at the mean, and in the interquartile range of gender diversity) will not see any significant difference between inclusion and lack of inclusion in terms of their creativity. We explore our predictions along varying levels of diversity, both for minimal and maximal inclusion, by keeping all controls constant at their means. Figure 4. shows the predictions at the developer firm level.

Across all three inclusion measures, and also their combined index, our predictions indicate that video game developer teams cannot increase their creativity at any level of gender diversity, if female developers are not included (their inclusion measures equal zero). In case of bonding and the combined index, this finding stays robust even if re-label 50 percent of unknown gendered team members as males. (See Materials and Methods, Impact of Gender Robustness on Modelling for more details, and Figure S4 for Point Estimates on relabeled data Figure S2, panel d in Supplementary Materials.) This suggests that adding "newbie" female team members without prior collaborative ties to male team members will not contribute to increased creativity in the first instance.

In contrast to predictions with minimal inclusion, we see that at maximal inclusion an increase in gender diversity leads to increase in creativity, measured by game distinctiveness. If a team moves from the lowest gender diversity in our dataset to the highest, while maintaining maximal inclusion, it can boost distinctiveness by 10% in the case of mixing, by 20% in the case of bonding, by 7% in the case of incorporating, and by 22% for combined inclusion.

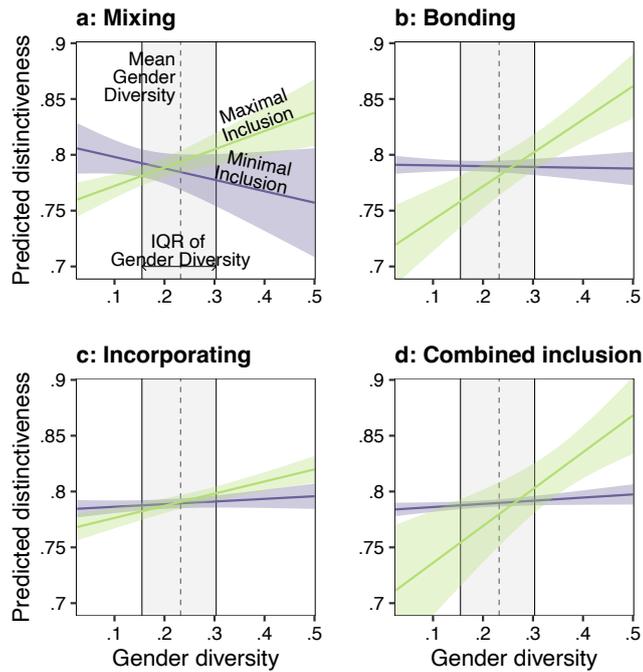

**Figure 4:** Predicted distinctiveness by gender diversity at minimal and maximal levels of inclusion. Panels show predictions along four measures of inclusion; shaded areas indicate 95% CI.

Our results indicate that a firm with only male developers would find it difficult to realize creativity benefits from adding female developers initially. Considering firms that include female developers for the first time after working with only male developers (there were 306 such firms in our dataset of the complete set of 1354 firms), we do not see any increase in creativity. The distinctiveness score of the first game with any females is even slightly (but not significantly) lower than the preceding game with males only. What our predictions indicate is that developer firms need to reach diversity that is higher than the top quartile to start seeing benefits from gender diversity, when female developers are also included in the team's network. Firms with low diversity would see inclusion decreasing the creativity of the team. This presents a barrier to seeing the incentives to boost diversity in video game developer firms. Nevertheless, several developer firms did successfully increase diversity and inclusion, and in the next section we attend to game histories of firms to understand the processes that can lead to higher levels of diversity and inclusion, despite the lack of early benefits.

*Firm-level processes that lead to diversity and inclusion*
How can firms boost diversity and inclusion? We turn to analyze histories of game developer firms, to understand if intervention in diversity, or intervention in inclusion is what leads to higher levels of diversity and inclusion at the end of their histories. In the first case, if boosting diversity is the key to advancing both diversity and inclusion, firms can add female developers to their teams, and then subsequently see an increase in inclusion, when female developers build ties to male developers in repeated game projects. In the second case, if firms can intervene by adding inclusion, the key is to hire subsets of developers with gender diversity *and* pre-existing ties between female and male team members. In this case, inclusion is not "home grown", but rather a function of clustered migration of individuals among firms. This process also occurs frequently, as was recently described as the "trojan horse" mechanism (Arvidsson, Collet, and Hedström 2021), driven by a sequence of clustered migration of individuals who have prior collaborative ties between them.

To capture the primary drivers of firm-level processes, we use transfer entropy, a measure that captures the amount of information values in one time series have about subsequent values of another time series (Barnett, Barrett, and Seth 2009; Schreiber 2000). Low transfer entropy means that a prior increase in the first process is not followed by an increase in the second process, while high transfer entropy means a strong directional coupling. We measure transfer entropy between two processes: the developer firm-level times series of diversity and inclusion, in both directions. We enter the resulting variables, transfer entropy $T_{D \to I}$ and $T_{I \to D}$ in an OLS model that predicts the final diversity and inclusion, and the trends of diversity and inclusion at the developer firm level. Figure 5 shows two examples, one where diversity predicts inclusion, and a second where inclusion predicts diversity.

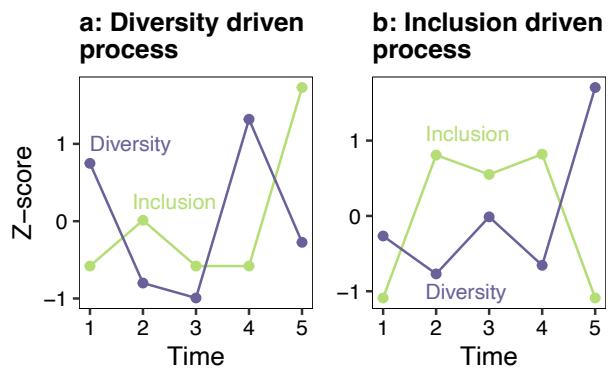

**Figure 5:** Examples of diversity driven ($T_{D \to I} = \max T_{D \to I}$) and inclusion driven ($T_{I \to D} = \max T_{I \to D}$) processes of length five.

Figure 6 shows point estimates predicting firm-level diversity and inclusion. We found that boosting diversity and inclusion seems to be a product of a diversity-driven process, where in a firm-level sequence of video game projects changes in diversity result in changes in inclusion. This process in practice can be conceptualized as the hiring of female developers regardless of their prior histories of collaboration with other team members, and subsequently adding inclusion by repeated collaborations between female and male developers – a form of "home grown" inclusion. The reverse direction of temporal influence between processes does not seem to lead to increased diversity or inclusion: When firms add inclusion first (and diversity is a result of this subsequently), we should expect no measurable advantage in increased diversity or inclusion. In practice such a process would mean hiring dyads of female and male developers with pre-existing collaboration ties, which we could label as "acquired inclusion".

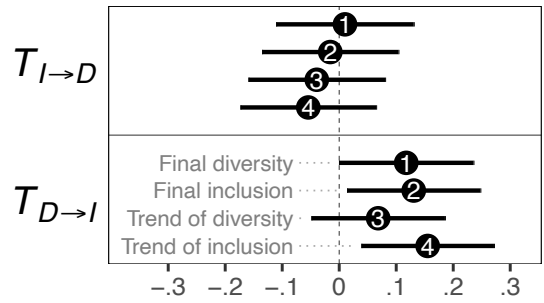

**Figure 6:** Point estimates for predicted diversity and inclusion (final value and trend) by the nature of firm-level processes: transfer entropy of inclusion to diversity, and transfer entropy from diversity to inclusion, when these transfer entropies are simultaneously entered in an OLS prediction. Horizontal axis shows predicted SD change in outcome when independent variables increase by one SD.

This indicates that developer firms should not hesitate to add novice female developers to their teams – even if they cannot expect immediate creativity benefits in the team with female developers without inclusion, as female developers would not yet have cross-gender collaborative ties. Female developers will accumulate collaborative ties, and thus achieve inclusion subsequently, and the team can expect to see a boost in creativity.

**Discussion**

As others have already found, gender diversity in itself is a predictor of creativity (Bear and Woolley 2011; Beede et al. 2011; Hofstra et al. 2020; Xie et al. 2020). However, when we take inclusion also into account, the main effect of gender diversity on creativity is not significant, suggesting that diverse collaborators in a team also need to be included to for the team to see creativity benefits. Gender diversity interacts with inclusion in a way that diversity without inclusion does not bring any advantages in creativity, regardless of the extent of diversity. The creative benefit of diversity increases only as much as inclusion increases.

Our results indicate that organizations should pay attention to inclusion as well, not only to diversity. There is a rich literature on inclusion stressing the importance of integrating employees and team members with diverse attributes (Chung et al. 2020, 2020; Ferdman and Deane 2014; Jansen et al. 2014; Mor-Barak and Cherin 1998), but systematic and large scale measurement tools for inclusion, diversity, and creativity were not developed in conjunction. We operationalize inclusion using the network of past collaboration, developing three diverse metrics that all support the same conclusion: gender diversity without inclusion does not lead to benefits in creativity.

At the same time, we also see evidence for why tokenism: the minimal presence of gender minority is not effective. When a team adds one female member, one can expect no creativity benefits. In fact, when we observe developer firms with exclusively male teams in their history adding a female developer for the first time, there is a slight decrease in creativity. This underscores prior findings about the limits of tokenism (Farh et al. 2020; Guldiken et al. 2019): when gender diversity is low, inclusion acts more as assimilation that silences the creative potential in diversity.

The process by which organizations include diverse collaborators is not indifferent, especially as a developer firm looking to increase diversity from zero, or very low levels would not see early benefits in creativity. We found evidence that it is more beneficial to first increase diversity, and then inclusion, rather than the other way around. Organizations should aim to recruit novice, unconnected female collaborators, and then increase inclusion by employing these novice female tema members in repeated projects. The alternative approach of recruiting diverse team members already with a history of cross-gender collaborations in prior projects does not lead to sustained increase in diversity and inclusion. Organizations need to add and include a relatively high proportion of female developers (about 23% - significantly higher than the industry average of 19%) to start seeing creativity benefits. This is a likely contributor to the sustained marginalization of female developers in the field, reinforcing beliefs in the benefits of male-skewed team composition.

**Limitations**

Limitations of our study chiefly relate to the definition and measurement of diversity and collaboration. Gender identity is not binary, however such personal information could be only analyzed if self-claimed gender identity is provided, therefore we could not incorporate non-binary gender into this study. We are also aware of the limitations and the potential biases of name-based inferring methods, such algorithms perform better on Western-names (Karimi et al. 2016). To account for the potential bias that the presence of unknowns within teams implies we performed robustness checks and found even if 50 percent of unknowns are male the positive interaction of gender diversity and inclusion persists in the case of combined index and bonding but mixing and incorporating are more sensitive to such bias.

Our measurement of past collaboration was restricted to collaborations within the population of games in our dataset, thus we have no data about past collaborations in game projects not in the database, or projects in other industries. To fully capture inclusion we would also need to have multiplex network data about communication and other relevant on-project relationships, as well as a subjective sense of acceptance. Our measure of diversity did not take dimensions beyond gender into account, while in collaborative settings complex intersectional diversity is at play.

**Materials and Methods**

*Data*
We collected data from the video game industry, relying on MobyGames.com[1] Our dataset contains 8,617 unique video games, with a list of each game's developer teams, critic's reviews, and stylistic elements such as genres, perspective (e.g., first-person shooter, role-playing) and the platforms it can be played on (e.g., PlayStation, Nintendo Switch, etc.). We also record each game's developer studio, publishing house, and the year of the first release. The video game industry has gone through a major change, with the rising popularity of mobile games in the early 2010s. As the industry became more competitive, and labour shortages hit the tech industry, companies stopped publishing the entire credit list of their project teams, probably to avoid offers sent to their employees from competitors. Therefore, our analysis covers games published between the 1980s to 2010.

---

[1] https://www.mobygames.com/, MobyGames is a website which catalogs video games via crowdsourcing. It covers 300 gaming platforms and over 230,000 games.

Since our database goes back to the very beginnings of the video game industry, we can infer everyone's full career path; connecting unique user accounts with the games they had worked on in a consecutive order. It allows us to create team-level weighted networks for each video game: two team members are connected based on how many times they had worked on the same game.

For our analysis we only considered games which were published between 1993 and 2009, and had less than 2000 connection among team members, had at least one female team member, and less than 50% of team members gender could have been inferred. We excluded all re-released and mobile games. Since gender diversity is a key interest of our study, we had to exclude all those video games from our analysis which did not list team members' full name and used only initials instead of first names. Our resulting database contains 4,011 video games. (For more details see *Table S1, S2 in* Supplementary Materials.)

*Gender Inferring*
Similarly, to film credits, Moby Games lists each team member's full name and task in the production (imaging, scripting, design, music, etc.). To infer team members' gender, we relied on developers' full names, and adopt a commonly used first-name based gender inferring method (Vedres and Vasarhelyi 2019). Name-based gender inferring methods has been criticized for treating gender as a binary category, and over-representing Western names (Karimi et al. 2016). The method that we selected is optimized for high precision, where names with high probability for being unisex are labelled as unknowns. Our gender inferring yielded 19 percent female, 63 percent male and 18 percent unknowns. (For further details on the accuracy of gender inferring see *Gender Inferring, Figure S1., Table S3 in* Supplementary Materials.)

**Quantitative Measures**

Dependent Variable: We measure *creativity* by adopting De Vann et al.'s distinctiveness metric, which compares the combination of each game's stylistic elements to all games released in the preceding five years and compute and average distance (1- cosine similarity) between them (de Vaan et al. 2015). Since we do not know the exact publish date of a very game, we did not compare games published within the same to avoid temporal aversion.

Cosine Distance $d_{i,j}$ is calculated 1) by comparing the vectors of stylistic elements of all game *i* with all other game *j,* the following:

$$d_{i,j} = 1 - \left[ \sum_{k=1}^{K} g_{ik}g_{jk} \middle/ \left(\sum_{k=1}^{K} g_{ik}^2\right)^{1/2} \left(\sum_{k=1}^{K} g_{jk}^2\right)^{1/2} \right]$$

Where $g_{ik}$ is *1/K* if a given stylistic element *k* was used in game *i* and 0 otherwise. Then the resulting similarity is subtracted from 1. 2) Finally we normalize these game-pair distances for all games (1,2,…,j) published in the proceeding 5 years, as the following:

$$Creativity = \sum_{j=1, j \neq i}^{N} d_{ij}/N$$

Independent Variables - Gender Diversity and Team cohesion metrics: Our core interest is how teams' gender diversity and inclusion predict creativity and success in the video game industry.

*Gender Diversity: Blau's index*
We use Blau's Heterogeneity Index as our measure of diversity. It is calculated as $H_B = 1 - \sum P_i^2$, where $P_i$ is the ratio of group members in category *i* (male or female). Therefore, the female-male ratio is 50-50 percent the Blau Index is 1, and when a team is composed only by one gender group is 0.

We measured inclusion in four ways by using network-based segregation metrics:

*Mixing as reversed assortativity*
Assortativity Coefficient developed by Mark Newman (Newman 2003) measures the similarity of connections in the graph with respect to the given attribute. It has been widely used to measure homophily in various (social) networks: such as sexual contacts and marriage matching (Girvan and Newman 2002; Newman 2003), demographics on Facebook (Traud, Mucha, and Porter 2012), book recommendation networks (Bucur 2019) or the research interests of scientists who follow each other on twitter (Ke, Ahn, and Sugimoto 2017). The Assortativity Coefficient, r is calculated as the Pearson correlation coefficient of degrees between pairs of nodes, formally $r = \frac{Tr(M) - \sum M^2}{1 - \sum M^2}$, where *M* is the mixing matrix (joint probability) of the two genders, and *Tr(M)* is the trace (sum of elements in the diagonal) of matrix *M*. *r=0* is where the network is perfectly disassortative, meaning that every edge connects a node to a different type, while *r=1* means perfect assortativity, when the network is fully segregated, such that nodes from type i do not connect to nodes to type j. We quantify reversed assortativity by subtracting it from one and normalizing it: $Mixing = |1 - r|/max(|1 - r|)$. Large values of Mixing mean high inclusion - team members are mixed by gender, and low values indicate gender segregation.

One of the beneficial attributes of assortativity coefficient while measuring segregation is that this metric is insensitive to the number of isolated nodes within the network (Bojanowski and Corten 2014). Because our collaboration networks are based on previous shared collaborations we have a higher number of isolated group members, which we should not consider while analyzing the network structure.

*Bonding as the ratio of weighted cross-gender ties*
More frequent shared project experience indicates more intense relationship among team members, which can be a proxy for higher inclusion. Women have been shown to strive and feel more included in workplaces where they could develop stronger ties (Timberlake 2005). Stronger ties were also shown to be beneficial to transfer complex knowledge (Hansen 1999; Nahapiet and Ghoshal 1997) and solve complex problems (de Montjoye et al. 2015). Therefore, our second metric quantifies gendered inclusion as the total number of times men and women worked together in previously divided by the total number of shared working experience of team: $Bonding = \sum W_{FM}/\sum_i W_i$, where $\sum W_{FM}$ is the sum of weights that connect different gender groups, and $\sum_i W_i$ is the sum of all weights within the network.

*Incorporating as the ratio of women in the graph center*
Our third inclusion metric captures how central women's position within the team network, specifically the ratio of women within the collaboration network's center. Network center is defined as the Jordan center of a graph, which is a set of nodes where eccentricity is equal to graphs' radius. The eccentricity $\epsilon(n)$ of a node $n$ measures how far a node is from the furthest node in the graph. Formally $\epsilon(n) = \max_{u \in N}(n, u)$. The radius $r$ of a graph is the minimum eccentricity of any node, formally $r = min(\epsilon(n)) = \min_{n \in N} \max_{u \in N}(n, u)$ To measure which team members belong to the center we used Python 3. Netwokx Center Distance Metric. Finally we take the natural logarithm of the ratio of women in the center to ensure a better distribution, therefore calculated as $Incorporating = log(\frac{N_{w \in C}}{N_w})$ , where $N_{w \in C}$ is the number of women in the center $N_w$ is the number of women in the team.

*Combined inclusion*
Our fourth measure of inclusion is the combination of the first three, as a product of the three measures of inclusion: mixing, bonding, and incorporating.

**Modelling Distinctiveness**

Our dependent variable that captures creativity - "distinctiveness" is a normally distributed variable we use OLS Regression models to model the impact diversity and inclusion on it. We ran separate models for each inclusion metric – mixing, bonding, incorporating and combined inclusion. To compare models we normalize, our dependent and independent variables by their minimum values as the following: $X_{norm} = (x - min_x)/std_x$

In each model we controlled for multiple attributes that can provide an alternative explanation for teams' creativity. *Team size* is measured as the number of team members involved in the game production. Larger teams are typically assembled by more established developer firms, therefore more likely to have bigger networks and employ more women. *Ratio of center:* Ratio of center measures how many percent of the team network belongs to the center. More flat organizations (with higher center) are more likely to be more democratic and allow minorities to share their ideas. *Number of Newbies*, measures the number of team members with no experience in game development (based on our database). We also counted for the *Number of star developers*, those who have been awarded a Game Developers Choice Award. *Game tenure* captures the experience level of a team, measured as the average number of games team members have produced prior to the year of production of the given game. *Single-Firm Production* Is a dummy variable, which is 1 if the publisher and the developer company is the same entity, otherwise 0. We controlled for the *platforms* the game was developed for, because certain genres and platforms can be more popular than others. We also controlled for temporal trends, with the t *year of release* and the *number of countries* the game was released at.

To account for factors that are not directly measurable on produced games' distinctiveness we provide three alternative modelling approaches: game-level analysis 1) with random effects for developer firm, 2) with fixed-effects for developer firm 3) and developer firm level aggregated analysis. *(See Figure S2)*.

*Impact of Gender Robustness on Modelling*

To account for the potential bias that inferred gender could introduce to results, we adopt various robustness checks. Although the precision of our name-based gender inferring method was nearly perfect for men and women, we accounted much lower precision for unknowns (50%). Although we excluded teams with more than 50% of unknowns from our analysis, there is still bias that team members with unknown gender can add to our results. Statistics on female representation in the video game industry indicate that most of the unknowns are more likely to be male. Therefore, we randomly select 25 or 50 percent of unknown gendered team members in each game and re-label them as males and re-calculate all diversity and inclusion metrics. We repeat this process 100 times and take the average of the resulting inclusion metrics for each game. (Figure S3, in SI shows the distribution of newly calculated gender diversity and inclusion metrics with 25 and 50 percent of relabelled data compared to original data). Then we rerun game-level OLS models to predict games' distinctiveness based on the 25 and 50 percent relabelled diversity and inclusion metrics. The interaction between gender diversity and bonding stays significant even if 50 percent of unknown gendered team members are labelled as males. Similarly to bonding, combined inclusion's interaction with gender diversity is robust to gender relabelling, while mixing and incorporating loose their significance if at least 25 percent of unknowns turn out to be male. (See the estimated coefficients of diversity and inclusion at Figure S4, SI *in* Supplementary Materials.)

*Time series analysis*

We have filtered our data to include games from developer firms that had at least four games in the dataset. This resulted in a dataset with 2418 games from 308 developer firms, filtered from the original dataset of 4011 games from 1354 firms. Distributions of key variables (creativity, diversity, and combined inclusion) in the filtered dataset did not differ from the full dataset (with Kolmogorov-Smirnov test p values of .99, .65, and .83 respectively), and the means of these variables were not significantly different either (with Wilcoxon rank sum test p-values of .57, .22, and .83 respectively). We recorded the diversity and combined inclusion scores for these games, and we calculated transfer entropy from diversity to inclusion, and from inclusion to diversity as $T_{X \rightarrow Y} = S(Y_t | Y_{t-1:t-L}) - S(Y_t | Y_{t-1:t-L}, X_{t-1:t-L})$, where $S(Y)$ is the Shannon entropy of $Y$.

Since time resolution for the publication date for games in annual, we had several games within a developer firm that were from the same year. For these games with tied dates we have used random sorting, and re-calculated transfer entropies. We used 500 random sortings of ties for all temporal sequences. We then calculated the mean transfer entropy scores of these 500 sequences for each developer firm game sequence.

# References

Adams, Renée B., and Patricia Funk. 2012. "Beyond the Glass Ceiling: Does Gender Matter?" *Management Science* 58(2):219–35. doi: 10.1287/mnsc.1110.1452.

Amabile, Teresa M. 1983. "The Meaning and Measurement of Creativity." Pp. 17–35 in *The social psychology of creativity*. Springer.

Amabile, Teresa M., Regina Conti, Heather Coon, Jeffrey Lazenby, and Michael Herron. 1996. "Assessing the Work Environment for Creativity." *Academy of Management Journal* 39(5):1154–84. doi: 10.5465/256995.

Aral, Sinan, and Marshall Van Alstyne. 2011. "The Diversity-Bandwidth Trade-Off." *American Journal of Sociology* 117(1):90–171. doi: 10.1086/661238.

Arvidsson, M., F. Collet, and P. Hedström. 2021. "The Trojan-Horse Mechanism: How Networks Reduce Gender Segregation." *Science Advances* 7(16):eabf6730. doi: 10.1126/sciadv.abf6730.

Bailey, Eric N., Kazunori Miyata, and Tetsuhiko Yoshida. 2021. "Gender Composition of Teams and Studios in Video Game Development." *Games and Culture* 16(1):42–64. doi: 10.1177/1555412019868381.

Barnett, Lionel, Adam B. Barrett, and Anil K. Seth. 2009. "Granger Causality and Transfer Entropy Are Equivalent for Gaussian Variables." *Phys. Rev. Lett.* 103(23):238701. doi: 10.1103/PhysRevLett.103.238701.

Bear, Julia B., and Anita Williams Woolley. 2011. "The Role of Gender in Team Collaboration and Performance." *Interdisciplinary Science Reviews* 36(2):146–53. doi: 10.1179/030801811X13013181961473.

Beckhusen, Julia. 2016. "Occupations in Information Technology." *US Census Bureau*.

Beede, David N., Tiffany A. Julian, David Langdon, George McKittrick, Beethika Khan, and Mark E. Doms. 2011. "Women in STEM: A Gender Gap to Innovation." *Economics and Statistics Administration Issue Brief* (04–11).

Begeny, C. T., M. K. Ryan, C. A. Moss-Racusin, and G. Ravetz. 2020. "In Some Professions, Women Have Become Well Represented, yet Gender Bias Persists—Perpetuated by Those Who Think It Is Not Happening." *Science Advances* 6(26):eaba7814. doi: 10.1126/sciadv.aba7814.

Bojanowski, Michał, and Rense Corten. 2014. "Measuring Segregation in Social Networks." *Social Networks* 39:14–32. doi: 10.1016/j.socnet.2014.04.001.

Brooke, Sian. 2019. "'Condescending, Rude, Assholes': Framing Gender and Hostility on Stack Overflow." Pp. 172–80 in *Proceedings of the Third Workshop on Abusive Language Online*. Florence, Italy: Association for Computational Linguistics.

Bucur, Doina. 2019. "Gender Homophily in Online Book Networks." *Information Sciences* 481:229–43. doi: 10.1016/j.ins.2019.01.003.

Carton, Andrew M., and Jonathon N. Cummings. 2012. "A Theory of Subgroups in Work Teams." *Academy of Management Review* 37(3):441–70. doi: 10.5465/amr.2009.0322.

Cheryan, Sapna, Sianna A. Ziegler, Amanda K. Montoya, and Lily Jiang. 2017. "Why Are Some STEM Fields More Gender Balanced than Others?" *Psychological Bulletin* 143(1):1–35. doi: 10.1037/bul0000052.

Chung, Beth G., Karen H. Ehrhart, Lynn M. Shore, Amy E. Randel, Michelle A. Dean, and Uma Kedharnath. 2020. "Work Group Inclusion: Test of a Scale and Model." *Group & Organization Management* 45(1):75–102. doi: 10.1177/1059601119839858.

Clark Blickenstaff, Jacob. 2005. "Women and Science Careers: Leaky Pipeline or Gender Filter?" *Gender and Education* 17(4):369–86. doi: 10.1080/09540250500145072.

Cohen, Susan G., and Diane E. Bailey. 1997. "What Makes Teams Work: Group Effectiveness Research from the Shop Floor to the Executive Suite." *Journal of Management* 23(3):239–90. doi: 10.1177/014920639702300303.

De Dreu, Carsten K. W., and Michael A. West. 2001. "Minority Dissent and Team Innovation: The Importance of Participation in Decision Making." *Journal of Applied Psychology*

86(6):1191–1201. doi: 10.1037/0021-9010.86.6.1191.

Farh, Crystal IC, Jo K. Oh, John R. Hollenbeck, Andrew Yu, Stephanie M. Lee, and Danielle D. King. 2020. "Token Female Voice Enactment in Traditionally Male-Dominated Teams: Facilitating Conditions and Consequences for Performance." *Academy of Management Journal* 63(3):832–56.

Ferdman, Bernardo M., and Barbara Deane, eds. 2014. *Diversity at Work: The Practice of Inclusion*. San Francisco, CA: Jossey-Bass, A Wiley Brand.

Girvan, M., and M. E. J. Newman. 2002. "Community Structure in Social and Biological Networks." *Proceedings of the National Academy of Sciences* 99(12):7821–26. doi: 10.1073/pnas.122653799.

Granovetter, Mark. 1973. "The Strength of Weak Ties." *American Journal of Sociology* 78(6):1360–80.

Guimera, R., B. Uzzi, J. Spiro, and L. A. Amaral. 2005. "Team Assembly Mechanisms Determine Collaboration Network Structure and Team Performance." *Science* 308(5722):697–702. doi: 10.1126/science.1106340.

Guldiken, Orhun, Mark R. Mallon, Stav Fainshmidt, William Q. Judge, and Cynthia E. Clark. 2019. "Beyond Tokenism: How Strategic Leaders Influence More Meaningful Gender Diversity on Boards of Directors." *Strategic Management Journal* 40(12):2024–46. doi: https://doi.org/10.1002/smj.3049.

Hansen, Morten T. 1999. "The Search-Transfer Problem: The Role of Weak Ties in Sharing Knowledge across Organization Subunits." *Administrative Science Quarterly* 44(1):82–111. doi: 10.2307/2667032.

Haverkamp, Andrea, Michelle Bothwell, Devlin Montfort, and Qwo-Li Driskill. 2021. "Calling for a Paradigm Shift in the Study of Gender in Engineering Education." *Studies in Engineering Education* 1(2):55. doi: 10.21061/see.34.

Hofstra, Bas, Vivek V. Kulkarni, Sebastian Munoz-Najar Galvez, Bryan He, Dan Jurafsky, and Daniel A. McFarland. 2020. "The Diversity–Innovation Paradox in Science." *Proceedings of the National Academy of Sciences* 117(17):9284–91. doi: 10.1073/pnas.1915378117.

Horwitz, Sujin K., and Irwin B. Horwitz. 2007. "The Effects of Team Diversity on Team Outcomes: A Meta-Analytic Review of Team Demography." *Journal of Management* 33(6):987–1015. doi: 10.1177/0149206307308587.

Jansen, Wiebren S., Sabine Otten, Karen I. van der Zee, and Lise Jans. 2014. "Inclusion: Conceptualization and Measurement: Inclusion: Conceptualization and Measurement." *European Journal of Social Psychology* 44(4):370–85. doi: 10.1002/ejsp.2011.

Jehn, Karen A. 1995. "A Multimethod Examination of the Benefits and Detriments of Intragroup Conflict." *Administrative Science Quarterly* 40(2):256. doi: 10.2307/2393638.

Joshi, Aparna, and Hyuntak Roh. 2009. "The Role of Context in Work Team Diversity Research: A Meta-Analytic Review." *Academy of Management Journal* 52(3):599–627. doi: 10.5465/AMJ.2009.41331491.

Karimi, Fariba, Claudia Wagner, Florian Lemmerich, Mohsen Jadidi, and Markus Strohmaier. 2016. "Inferring Gender from Names on the Web: A Comparative Evaluation of Gender Detection Methods." *Proceedings of the 25th International Conference Companion on World Wide Web - WWW '16 Companion* 53–54. doi: 10.1145/2872518.2889385.

Ke, Qing, Yong-Yeol Ahn, and Cassidy R. Sugimoto. 2017. "A Systematic Identification and Analysis of Scientists on Twitter" edited by L. Bornmann. *PLOS ONE* 12(4):e0175368. doi: 10.1371/journal.pone.0175368.

van Knippenberg, Daan, Carsten K. W. De Dreu, and Astrid C. Homan. 2004. "Work Group Diversity and Group Performance: An Integrative Model and Research Agenda." *Journal of Applied Psychology* 89(6):1008–22. doi: 10.1037/0021-9010.89.6.1008.

Kooistra, Jelle. 2019. "Newzoo's Trends to Watch in 2019." Retrieved February 8, 2022 (https://newzoo.com/insights/articles/newzoos-trends-to-watch-in-2019/).

Kozlowski, Steve W. J., and Bradford S. Bell. 2013. "Work Groups and Teams in Organizations." Pp. 412–69 in *Handbook of Psychology, Volume 12: Industrial and Organizational*

*Psychology*, edited by I. B. Weiner, N. W. Schmitt, and S. Highhouse. John Wiley & Sons, Inc.

Levi, Daniel. 2017. *Group Dynamics for Teams*. 5th edition. Los Angeles: SAGE.

Lutter, Mark. 2015. "Do Women Suffer from Network Closure? The Moderating Effect of Social Capital on Gender Inequality in a Project-Based Labor Market, 1929 of 2010." *American Sociological Review* 80(2):329–58. doi: 10.1177/0003122414568788.

Milliken, Frances J., and Luis L. Martins. 1996. "Searching for Common Threads: Understanding the Multiple Effects of Diversity in Organizational Groups." *Academy of Management Review* 21(2):402–33. doi: 10.5465/amr.1996.9605060217.

de Montjoye, Yves-Alexandre, Arkadiusz Stopczynski, Erez Shmueli, Alex Pentland, and Sune Lehmann. 2015. "The Strength of the Strongest Ties in Collaborative Problem Solving." *Scientific Reports* 4(1):5277. doi: 10.1038/srep05277.

Mor Barak, Michàlle E., Erica Leeanne Lizano, Ahraemi Kim, Lei Duan, Min-Kyoung Rhee, Hsin-Yi Hsiao, and Kimberly C. Brimhall. 2016. "The Promise of Diversity Management for Climate of Inclusion: A State-of-the-Art Review and Meta-Analysis." *Human Service Organizations: Management, Leadership & Governance* 40(4):305–33. doi: 10.1080/23303131.2016.1138915.

Mor-Barak, Michal E., and David A. Cherin. 1998. "A Tool to Expand Organizational Understanding of Workforce Diversity: Exploring a Measure of Inclusion-Exclusion." *Administration in Social Work* 22(1):47–64. doi: 10.1300/J147v22n01_04.

Nahapiet, Janine, and Sumantra Ghoshal. 1997. "Social Capital, Intellectual Capital and the Creation of Value in Firms." *Academy of Management Proceedings* 1997(1):35–39. doi: 10.5465/ambpp.1997.4980592.

Newman, M. E. J. 2001. "The Structure of Scientific Collaboration Networks." *Proceedings of the National Academy of Sciences* 98(2):404–9. doi: 10.1073/pnas.98.2.404.

Newman, M. E. J. 2002. "Assortative Mixing in Networks." *Physical Review Letters* 89(20):208701. doi: 10.1103/PhysRevLett.89.208701.

Newman, M. E. J. 2003. "Mixing Patterns in Networks." *Physical Review E* 67(2):026126. doi: 10.1103/PhysRevE.67.026126.

Nishii, Lisa H., and Jack A. Goncalo. 2008. "Demographic Faultlines and Creativity in Diverse Groups." Pp. 1–26 in *Research on Managing Groups and Teams*. Vol. 11. Bingley: Emerald (MCB UP ).

Northcraft, Gregory B., Jeffrey T. Polzer, Margaret A. Neale, and Roderick M. Kramer. 1995. "Diversity, Social Identity, and Performance: Emergent Social Dynamics in Cross-Functional Teams." in *Diversity in work teams: Research paradigms for a changing workplace.*, edited by S. E. Jackson and M. Ruderman. American Psychological Association.

Pearsall, Matthew J., Aleksander P. J. Ellis, and Joel M. Evans. 2008. "Unlocking the Effects of Gender Faultlines on Team Creativity: Is Activation the Key?" *Journal of Applied Psychology* 93(1):225–34. doi: 10.1037/0021-9010.93.1.225.

Pelled, Lisa Hope. 1996. "Demographic Diversity, Conflict, and Work Group Outcomes: An Intervening Process Theory." *Organization Science* 7(6):615–31. doi: 10.1287/orsc.7.6.615.

Phillips, Damon J. 2011. "Jazz and the Disconnected: City Structural Disconnectedness and the Emergence of a Jazz Canon, 1897-1933." *American Journal of Sociology* 117(2):420–83. doi: 10.1086/661757.

Reagans, Ray, and Ezra W. Zuckerman. 2001. "Networks, Diversity, and Productivity: The Social Capital of Corporate R&D Teams." *Organization Science* 12(4):502–17. doi: 10.1287/orsc.12.4.502.10637.

Rosener, Judy B. 2011. "Ways Women Lead." Pp. 19–29 in *Leadership, Gender, and Organization*. Vol. 27, *Issues in Business Ethics*, edited by P. Werhane and M. Painter-Morland. Dordrecht: Springer Netherlands.

Schreiber, Thomas. 2000. "Measuring Information Transfer." *Physical Review Letters* 85(2):461–64. doi: 10.1103/PhysRevLett.85.461.

Shore, Lynn M., Amy E. Randel, Beth G. Chung, Michelle A. Dean, Karen Holcombe Ehrhart, and Gangaram Singh. 2011. "Inclusion and Diversity in Work Groups: A Review and

Model for Future Research." *Journal of Management* 37(4):1262–89. doi: 10.1177/0149206310385943.

Timberlake, Sharon. 2005. "Social Capital and Gender in the Workplace." *Journal of Management Development* 24(1):34–44. doi: 10.1108/02621710510572335.

Traud, Amanda L., Peter J. Mucha, and Mason A. Porter. 2012. "Social Structure of Facebook Networks." *Physica A: Statistical Mechanics and Its Applications* 391(16):4165–80. doi: 10.1016/j.physa.2011.12.021.

United States Department of Labor. 2015. *Women's Bureau (WB) - Computer and Information Technology Occupations*.

Uzzi, Brian, and Jarrett Spiro. 2005. "Collaboration and Creativity: The Small World Problem." *American Journal of Sociology* 111(2):447–504. doi: 10.1086/432782.

de Vaan, Mathijs, Balazs Vedres, and David Stark. 2015. "Game Changer: The Topology of Creativity." *American Journal of Sociology* 120(4):1144–94. doi: 10.1086/681213.

Vedres, Balazs, and David Stark. 2010. "Structural Folds: Generative Disruption in Overlapping Groups." *American Journal of Sociology* 115(4):1150–90. doi: 10.1086/649497.

Vedres, Balazs, and Orsolya Vasarhelyi. 2019. "Gendered Behavior as a Disadvantage in Open Source Software Development." *EPJ Data Science* 8(1). doi: 10.1140/epjds/s13688-019-0202-z.

Webber, Sheila Simsarian, and Lisa M. Donahue. 2001. "Impact of Highly and Less Job-Related Diversity on Work Group Cohesion and Performance: A Meta-Analysis." *Journal of Management* 27(2):141–62. doi: 10.1177/014920630102700202.

Williams, Dmitri, Nicole Martins, Mia Consalvo, and James D. Ivory. 2009. "The Virtual Census: Representations of Gender, Race and Age in Video Games." *New Media & Society* 11(5):815–34. doi: 10.1177/1461444809105354.

Woodman, Richard W., John E. Sawyer, and Ricky W. Griffin. 1993. "Toward a Theory of Organizational Creativity." *Academy of Management Review* 18(2):293–321. doi: 10.5465/amr.1993.3997517.

Woolley, Anita Williams, Christopher F. Chabris, Alex Pentland, Nada Hashmi, and Thomas W. Malone. 2010. "Evidence for a Collective Intelligence Factor in the Performance of Human Groups." *Science* 330(6004):686–88. doi: 10.1126/science.1193147.

Wuchty, Stefan, Benjamin F. Jones, and Brian Uzzi. 2007. "The Increasing Dominance of Teams in Production of Knowledge." *Science* 316(5827):1036–39. doi: 10.1126/science.1136099.

Xie, Luqun, Jieyu Zhou, Qingqing Zong, and Qian Lu. 2020. "Gender Diversity in R&D Teams and Innovation Efficiency: Role of the Innovation Context." *Research Policy* 49(1):103885. doi: 10.1016/j.respol.2019.103885.

**Supplementary Text**

*Gender Inferring*

To quantify the error in our algorithm we randomly selected 200-200 team members whose gender was inferred as male, female and unknown and manually inferred their gender. Since these people re professionals, we look-up their profiles on LinkedIn and labelled them based on their pictures, bio, and recommendations (if the text used gendered language aka, He or She). If we could not find any LinkedIn user with the same name, or based on the picture or the text it was not possible to decide whether the given individual is a male or female, we labeled them as unknown.

Furthermore, we ran another widely used gender inferring method on our database gender-guesser[2] to compare how well our method performed. Figure S1 shows the accuracy of the used gender inferring method and the Gender Guesser Python package in comparison with the manually created baseline. Precision measures how many contributors were assigned to the correct gender label according to our baseline. Recall measures how many of the contributors were correctly identified. F-score takes the harmonic average of these two metrics. Our method has a higher precision for females and males, but lower recall for male. Table S3. Shows that our method is also better identifying unknowns than the Gender-Guesser default python package, but worse in identifying male contributors.

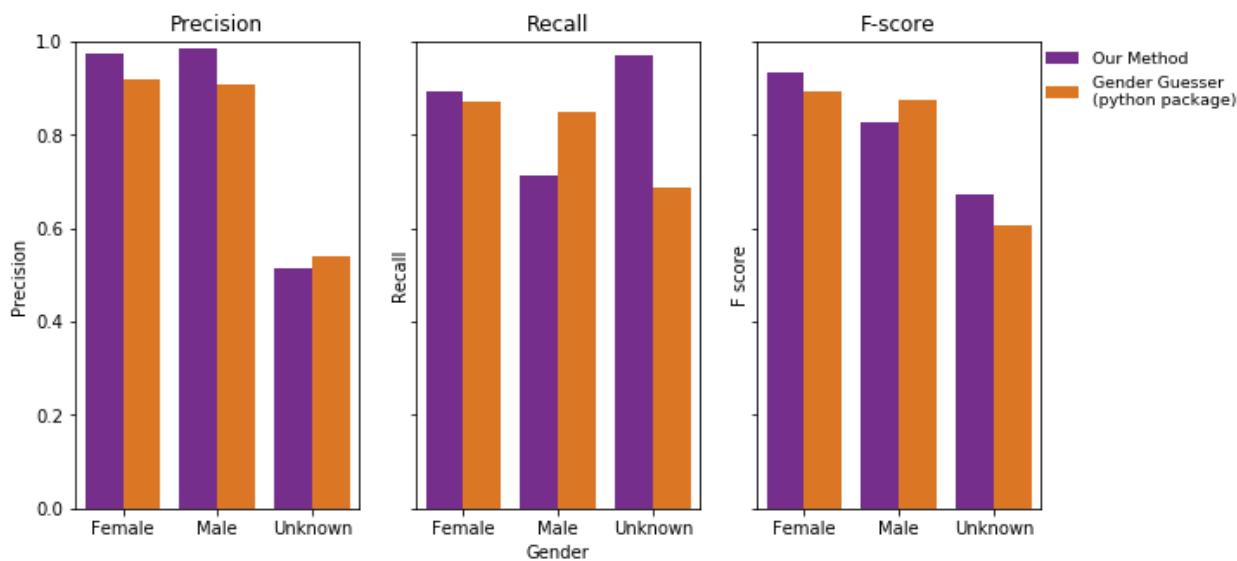

**Figure S1: Gender Inferring Accuracy.** Precision, Recall and F-score of our gender inferring method and the Gender Guesser Python package in comparison with the manually created baseline. Precision measures how many contributors were assigned to the correct gender label according to our manually created baseline. Recall measures how many of the contributors were correctly identified. F-score takes the harmonic average of these two metrics

---

[2] https://pypi.org/project/gender-guesser/

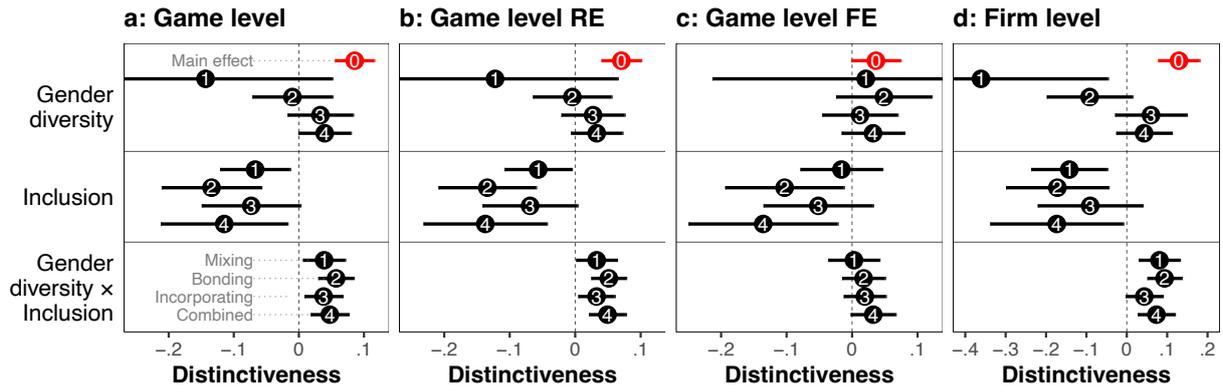

**Figure S2: Point estimates of distinctiveness in four model specifications.** Point estimates of distinctiveness with 95% CI for gender diversity, four variables of inclusion, and their interactions with gender diversity. Markers are numbered according to OLS models; coefficients are for one SD change in distinctiveness because of one SD change in independent variables. Panel a) shows the point estimates of different inclusion models based on the baseline game-level OLS models shown in the manuscript. Panel b) shows estimates for game-level OLS models with Random Effects to estimate the effect of game specific characteristics and Panel c) with Fixed-Effects for firm belonging to account for firm-level specific effects. Point estimates from models visualized in Panel d) are coming from firm-level aggregated data, where each variable is the average of all games produced by a given firm.

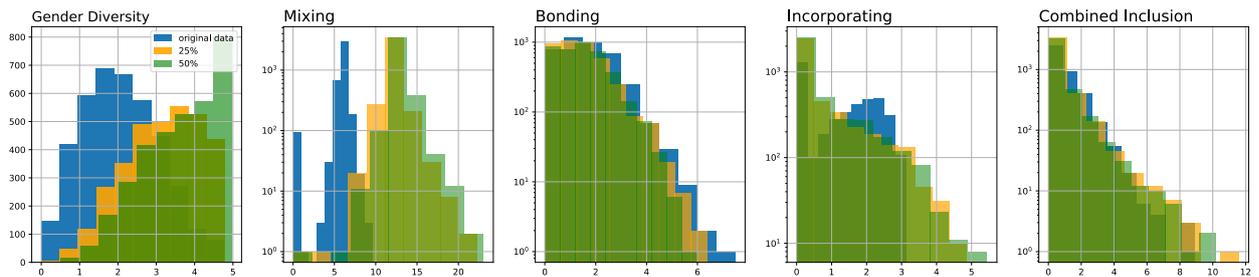

**Figure S3: Distribution of gender swapped diversity and inclusion metrics.** To assess the bias that name-based gender inferring method can introduce to our independent variables we randomly re-labeled 25 and 50 percent of unknown gendered team members to male in each game, and calculated Gender Diversity, Mixing, Bonding, Incorporating and Combined Inclusion. We repeated this process 100 times for each game and calculated the average of the resulting diversity and inclusion metrics. Blue histogram shows the original distribution, orange distribution is based on 25 percent re-labeled data, and green is based on 50 percent. Distributions indicate that Mixing and Incorporating are more sensitive to re-labelling than Bonding and the Combined Index.

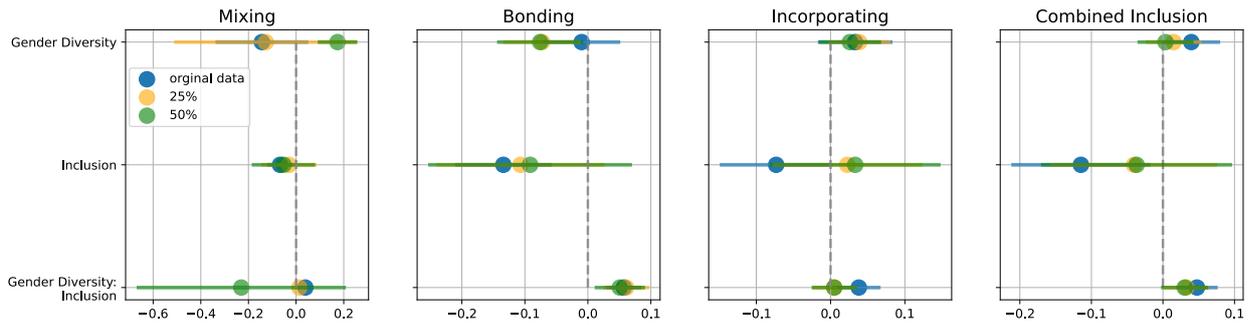

**Figure S4: Point estimates of distinctiveness based on OLS models ran on relabeled gender data.** To assess the impact of unknowns in project teams, we re-calculated our diversity and inclusion metrics 100 times by randomly relabeling 25 and 50 of unknown gendered team members to male. Blue indicates original point estimates of distinctiveness with 95% CI for gender diversity, yellow 25 percent of unknowns relabeled to male, and green 50.

| Number of games | 8,617 |
|---|---|
| Years | 1993-2009 |
| Number of developers | 630,420 |
| Women | 119,826 (19%) |
| Men | 397,520 (63%) |
| Unknowns | 113,074 (18%) |

**Table S1: Descriptive Statistics of original data collected from mobygames.com** We collected data from the video game industry, relying on MobyGames.com. Our dataset contains 8,617 unique video games, with a list of each game's developer teams, critic's reviews, and stylistic elements such as genres, perspective (e.g., first-person shooter, role-playing) and the platforms it can be played on (e.g., PlayStation, Nintendo Switch, etc.). We also record each game's developer studio, publishing house, and the year of the first release.

| **Filtering Criteria** | **N** |
|---|---|
| All games | 8,617 |
| Published between 1993 and 2009 | 7,931 |
| 1<number of edges in the network <2000 | 4,771 |
| Ratio of unknowns in team < 0.5 | 4,654 |
| Number of women in team network >=1 | **4,011** |

**Table S2: Applied filtering criteria on our game dataset.** For our analysis we only considered games which were published between 1993 and 2009, and had less than 2000 connection among team members, had at least one female team member, and less than 50% of team members gender could have been inferred. We excluded all re-released and mobile games. Since gender diversity is a key interest of our study, we had to exclude all those video games from our analysis which did not list team members' full name and used only initials instead of first names. Our resulting database contains 4,011 video games.

|  |  | Gender Guesser | | |
|---|---|---|---|---|
|  |  | Female | Male | Unknown |
| Our Method | Female | .895 | .040 | .065 |
|  | Male | .005 | .961 | .035 |
|  | Unknown | .135 | .292 | .575 |

**Table S3: Ratio of matching gender categories by Our method and Gender-Guesser.** Female contributors matched in 89.5%, and male contributors 96% of our sample.

| Model | Key Variables | SSR | SS_DIFF | F | P |
|---|---|---|---|---|---|
| 1 | Gender diversity **(compared to controls)** | 3,573 | 28,031 | 31,234 | .000 |
| 2 | Gender diversity+Mixing | 3,567 | 5,677 | 3,166 | .042 |
| 3 | Gender diversity+Bonding | 3,558 | 15,050 | 8,416 | .000 |
| 4 | Gender diversity+Incorporating | 3,567 | 5,915 | 3,299 | .037 |
| 5 | Gender diversity+Combined Inclusion | 3,561 | 11,359 | 6,345 | .002 |

**Table S4:** Game-level models efficiency gain and significance tests by models Model 1) F-test is calculated based on baseline model that contains only control variables. Models 2 to 5 are compared with Model 1. Significant F-test (P < 0.05) means that the analysed model explains the variance of creativity statistically better than the baseline model. SSR: Sum of squares of residuals in models, SS_DIFF: Difference in sum of squares compared to baseline models, F: value of statistic used to compare SSR of two models.